\RequirePackage{fix-cm}
\documentclass[smallextended]{svjour3}       % onecolumn (second format)
\smartqed  % flush right qed marks, e.g. at end of proof
\usepackage{graphicx}
\usepackage{latexsym}
\def\bfB {{\bf B}}
\usepackage{amssymb}

\begin{document}

\title{Ideal MHD instabilities for coronal mass ejections
%\thanks{Grants or other notes
%about the article that should go on the front page should be
%placed here. General acknowledgments should be placed at the end of the article.}
}
\subtitle{Interacting current channels and particle acceleration}

\titlerunning{MHD instabilities for CMEs}        % if too long for running head

\author{Rony~Keppens$^{*}$  \and Yang~Guo \and Kirit~Makwana \and
         Zhixing~Mei \and  Bart~Ripperda \and Chun~Xia \and  Xiaozhou~Zhao
}

%\authorrunning{Short form of author list} % if too long for running head

\institute{{\em $ ^{*}$Corresponding author:} R. Keppens \at
               Centre for mathematical Plasma Astrophysics, KU Leuven, 3001 Leuven, Belgium\\
              %Tel.: +32-16327001 
              %Fax: +32-16327998
              \email{rony.keppens@kuleuven.be}           %  \\
%             \emph{Present address:} of F. Author  %  if needed
\and Y. Guo \at School of Astronomy and Space Science, Nanjing University, Nanjing 210023, PR China \\
\and K. Makwana \at Deutsches Elektronen-Synchrotron DESY, Platanenallee 6, D-15738 Zeuthen, Germany \\
\and Z. Mei \at Yunnan Observatories, Chinese Academy of Sciences, Kunming, PR China \\
\and B. Ripperda \at Institut fur Theoretische Physik, Max-von-Laue-Str. 1, D-60438 Frankfurt, Germany  \\         
\and C. Xia \at Yunnan University, School of Physics and Astronomy, Kunming, PR China \\
\and X. Zhao \at Purple Mountain Observatory, Nanjing, PR China
}

\date{Received: date / Accepted: date}
% The correct dates will be entered by the editor

\maketitle

\begin{abstract}
We review and discuss insights on ideal magnetohydrodynamic (MHD) instabilities that can play a role in destabilizing solar coronal flux rope structures. For single flux ropes, failed or actual eruptions may result from internal or external kink evolutions, or from torus unstable configurations. We highlight recent findings from 3D magnetic field reconstructions and simulations where kink and torus instabilities play a prominent role. 

For interacting current systems, we critically discuss different routes to coronal dynamics and global eruptions, due to current channel coalescence or to tilt-kink scenarios. These scenarios involve the presence of two nearby current channels and are clearly distinct from the popular kink or torus instability. Since the solar corona is pervaded with myriads of magnetic loops -- creating interacting flux ropes typified by parallel or antiparallel current channels as exemplified in various recent observational studies -- coalescence or tilt-kink evolutions must be very common for destabilizing adjacent flux rope systems. Since they also evolve on ideal MHD timescales,  they may well drive many sympathetic eruptions witnessed in the solar corona. Moreover, they necessarily lead to thin current sheets that are liable to reconnection. We review findings from 2D and 3D MHD simulations for tilt and coalescence evolutions, as well as on particle acceleration aspects derived from computed charged particle motions in tilt-kink disruptions and coalescing flux ropes. The latter were recently studied in two-way coupled kinetic-fluid models, where the complete phase-space information of reconnection is incorporated.
% \PACS{PACS code1 \and PACS code2 \and more}
% \subclass{MSC code1 \and MSC code2 \and more}
\end{abstract}

\section{Introduction: forming flux ropes}\label{s-intsro}

\begin{figure*}
\centerline{\includegraphics[width=\textwidth]{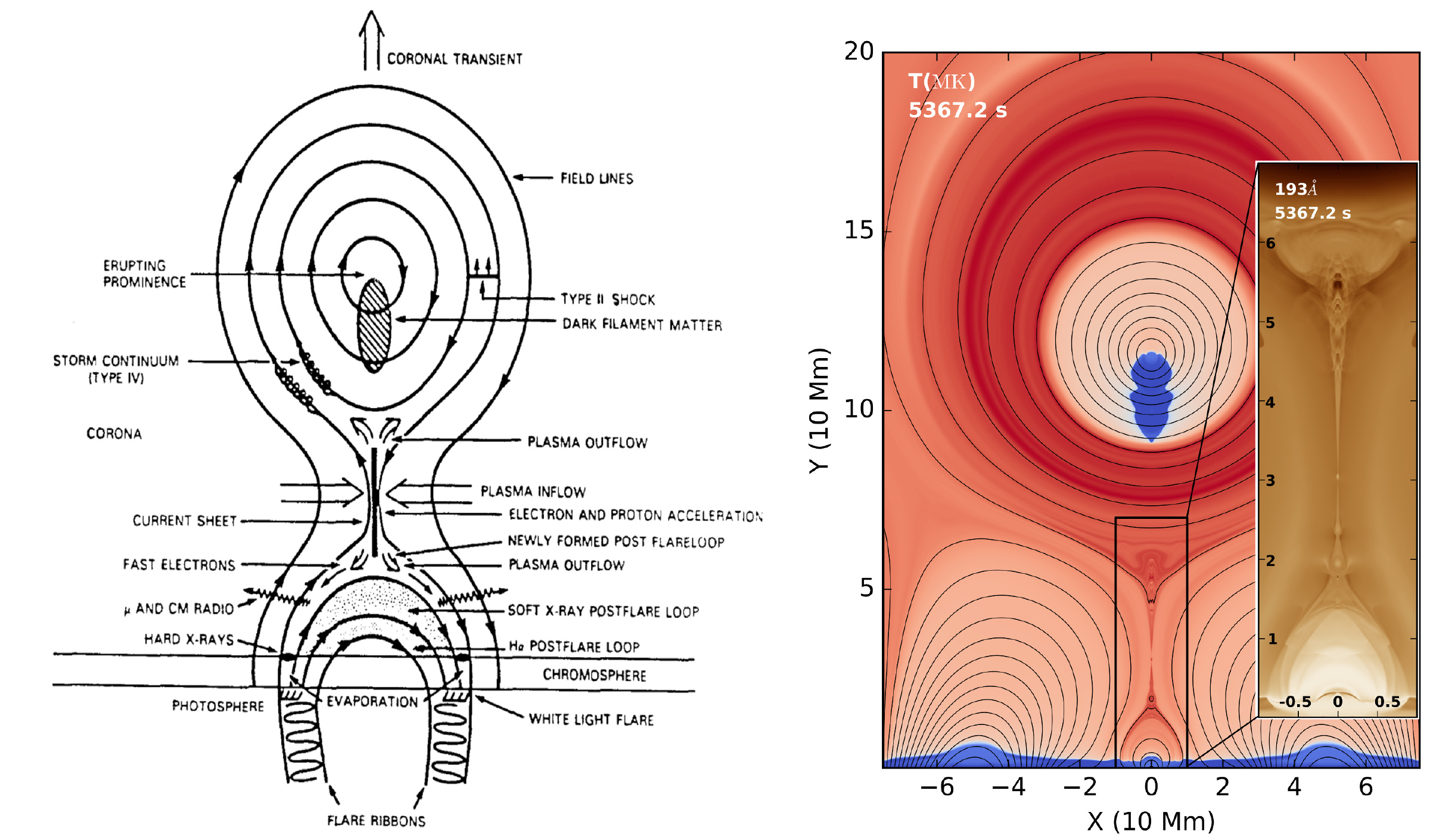}}
\caption{A cartoon (left) of the standard solar flare model ingredients, from~\cite{Martens1989}, contrasted with an MHD simulation (right) in a realistically stratified solar atmosphere, from~\cite{Zhaoetal2017,Zhaoetal2019}. The right panel inset zooms in on the current sheet dynamics, as seen in EUV emission. An erupting flux rope, with embedded prominence matter, shows reconnection and post-flare loops underneath.}\label{f-FR}
\end{figure*}

In what is currently referred to as `standard solar flare model', magnetic reconnection underneath an erupting flux rope plays a central role. This standard model has a long history, with early contributions by~\cite{CSHKP1,CSHKP2,CSHKP3,CSHKP4} (hence, CSHKP), and a historic overview can be found in the review by~\cite{Shibata2011}. Figure~\ref{f-FR}, taken from~\cite{Martens1989} (left panel) and \cite{Zhaoetal2017,Zhaoetal2019} (right panel), contrasts an early cartoon view with a modern simulation result of the eruption process. The cartoon identifies various observationally established features, such as the prominence matter co-erupting with the magnetic flux rope that supports it, or the formation of post-flare loops underneath the reconnection site. Post-flare loops are filled with hot evaporated material, emitting thermally in Extreme Ultra-Violet (EUV) and soft X-rays. The simulated view at the right shows a temperature view on a virtual eruption, while the zoomed inset gives synthetic EUV emission from the low-lying post-flare loops, but also from dynamics in the reconnection layer. This inset shows chaotic island formation within the current sheet, accompanied by the dynamic appearance of multiple X-points, each in turn locally demonstrating the typical Petschek-type standing slow shock structures~\cite{Petschek1964}. To simulate this properly, extreme (adaptive) numerical resolution was adopted: the minimal cell size corresponds to about 24 km, rivaling current and forthcoming\footnote{E.g., the forthcoming DKIST 4-meter solar telescope targets 0.1 arc-second resolution, or about 70 km in the solar photosphere. The Visual Tunable Filtergraph (VTF) instrument on DKIST may achieve 20 km spatial resolution at 520 nm. The balloon-borne SUNRISE missions achieved 50 km resolutions.} observational limits. Especially new in this simulation~\cite{Zhaoetal2017,Zhaoetal2019} is the self-consistent formation of an embedded prominence within the erupting flux rope, by scooping up chromospheric material: an initial arcade field got deformed by converging foot point motions into the typical solar flare model setup. Prominences that form in this way are thus directly levitated from chromospheric layers. Prominence formation by levitation is just one of several routes now known to lead to flux-rope embedded filaments. A lot of work has been done on prominences, both theoretically and through observations, and this can be found in various reviews~\cite{Gibson2018,Parenti2014}. In this review, we will rather focus on MHD stability aspects of magnetic flux ropes.

Although Fig.~\ref{f-FR} shows a 2D view on flux rope formation, the same process works equally in true 3D settings. How controlled (in particular, converging) foot point motions deform arcades to stable, 3D flux rope structures with overlying arcade fields was first demonstrated using unstratified, essentially zero-$\beta$ (where $\beta$ quantifies the ratio of thermal to magnetic pressure) models~\cite{Amari1999} that follow a series of force-free configurations. More recently, this was accomplished in an isothermal, stratified magnetohydrodynamic (MHD) simulation~\cite{Xia2014} where pressure and gravity both play a role. In the present review, special attention will be paid to aspects that require finite plasma $\beta$ effects. The latter is obviously needed to address solar filament aspects, which can also form in-situ by condensing evaporated, trapped plasma within the flux rope, to settle against gravity on dipped field line portions~\cite{Xiaetal2014,Xia2016}. The incorporation of gravity leads to roughly elliptic cross-sectional flux contours, much akin to the 2D views presented in Fig.~\ref{f-FR}. Hence, from a theoretical perspective, flux rope structures are a natural outcome when magnetic arcades reshuffle by convective motions. The twist within the rope relates to the current distribution, and magnetic flux ropes represent current-carrying field concentrations. In this review, we will primarily focus attention to ideal MHD mechanisms invoked for explaining solar eruptions, involving a single or multiple flux ropes. This leaves out models where reconnection is a primary means to drive coronal mass ejections, such as associated with the breakout model~\cite{Antiochos1999}, or tether-cutting descriptions~\cite{Moore1980}, while we will also not discuss coronal mass ejections (CMEs) initiation driven by means of, e.g., flux cancellation or emergence mechanisms. A complementary review on observational aspects of CMEs and the various triggering mechanisms that have been proposed, can be found in the paper by~\cite{Chen2011}.

\section{Flux rope stability: the usual suspects}

Whether current-carrying flux ropes (and their embedded filaments) ultimately erupt as coronal mass ejections is entirely dictated by stability considerations. For low-$\beta$ solar coronal conditions, the single-fluid MHD viewpoint is perfectly adequate, although filaments themselves represent partially ionized structures. As in man-made low-$\beta$ conditions in laboratory fusion devices, ideal MHD disruptions are the most violent ones, and mitigating their role has been a major achievement towards controlled nuclear fusion~\cite{Hansbook3}. Virtually all know-how on the delicate interplay between equilibrium and linear stability considerations has matured in laboratory context, and has gradually been transferred and adapted to solar coronal setups. We will first recall findings on both kink and torus instabilities, which we will from hereon refer to as `the usual suspects'. 
\subsection{Internal and external kinks for diffuse plasma columns}

In cylindrical models of flux rope structures, the twist can be quantified by the radial variation of the safety factor $q(r)=rB_z(r)/R_0B_\theta(r)$, where $2\pi R_0$ denotes the flux tube length, and $a$ its radius. Together, they define the inverse aspect ratio $\epsilon=a/R_0$, which links the cylinder to a slender `straight' tokamak when $\epsilon\ll 1$. The Kruskal-Shafranov limit, then expressed as $q(a)>1$, implies that a stabilizing axial field $B_z$, in combination with the finite length, can avoid the otherwise runaway process of an external kink instability \cite{Kruskal1958,Shafranov1956}. The external kink is in its purest form for a $z$-pinch configuration where a constant axial current $I_z$ coincides with the plasma region $r\leq a$, surrounded by a vacuum-wall setup. The kink is associated with a poloidal mode number $m=1$ in the $\exp(i m\theta+i k z)$ Fourier decomposition for linear perturbations, and this mode displaces the central axis helically. The external nature of the mode relates to the fact that the surrounding vacuum magnetic field variation is also perturbed in the process.

Internal kinks are associated with the existence of singular radial positions, where $q(r_s)=1$ for an $0<r_s<a$. In a torus-shaped low-$\beta$ tokamak, ensuring $q(0)>1$ while $q(a)>1$ guarantees stability to both external and internal kink modes. For cylindrical (1D) diffuse linear plasma columns, the theory of MHD stability is well established and can be found in textbook descriptions~\cite{Hansbook,Hansbook3}.  Details of the internal magnetic field variation determine many analytic stability results, while the pressure variation also appears in the established Suydam criterion for interchange/fluting modes \cite{Suydam1958}. The latter are connected to clustering sequences of discrete modes towards the edges of the MHD continua (frequency ranges $\omega(r)$ of slow and Alfv\'en frequencies), which extend to marginal frequency $\omega^2=0$ when $mB_\theta/r+k B_z$ vanishes internal to the plasma column. These locations are then rational-valued $q$ surfaces, where the field lines close upon themselves after a finite number of poloidal and toroidal (axial) turns.

Coronal flux ropes are also well-known to deviate from uniformly twisted configurations, but the detailed variation of $q(r)$ is hard to quantify precisely from observations. \cite{Liu2016} reported on nonlinear force-free field (NLFFF) models for a sequence of flares in AR 11817, identifying a fieldline specific twist number $T_w\equiv \int_L \nabla \times \bfB \cdot \bfB/(4\pi B^2) \, dl$,  which is readily integrated along each individual field line and relates well to the actual twist profile about its axis. \cite{Wang2017} identified how a complete flux rope structure was formed during eruption, typified by a varying-twist profile. Further out in the heliosphere, the magnetic topology of interplanetary (magnetic cloud) flux ropes can be studied in detail, and~\cite{Hu2014} found many interplanetary CMEs that deviate from uniform twist, or linear force-free states.

\subsection{Kink instability and line-tying}

The kink instability in solar coronal flux ropes is actually modified from the infinite cylinder, or from the straight tokamak, configuration. The reason is the stabilizing effect of line-tying, as both foot points of any typical coronal loop anchor firmly in the dense solar photosphere.  In~\cite{Hood1979}, it was shown analytically that a line-tied version of the cylinder setup is kink unstable only when the twist in the flux tube, related to the $q$-profile through $\Phi=2\pi/q$, exceeds a critical value. This critical twist value again depends on the equilibrium details, and a value of $\Phi_c=3.3\pi$ was found for the special case of a force-free, uniform twist equilibrium. This is indeed increased compared to the value $2\pi$ corresponding to the original Kruskal-Shafranov limit. Once more, the actual critical twist value turns out to be rather sensitive to equilibrium details, in particular to the $q(r)$ variation, the plasma-beta $\beta=2\mu_0p/B^2$ and the loop aspect ratio, i.e. the $\epsilon$ parameter. The analysis of~\cite{Hood1979} did not distinguish between the various types of kink modes mentioned earlier, as it considered a radially infinitely extended plasma column, i.e. a configuration where the `wall' is moved off to infinity, and where there is no vacuum region surrounding the plasma. The analysis did incorporate pressure effects and used the energy principle to assert instability against $m=1$ modes displacing the axis, for specific radial equilibrium configurations, i.e. pressure $p(r)$ and magnetic field components $B_z(r)$ and $B_\theta(r)$ ensuring force balance. 

\subsection{Torus instability}
In~\cite{Kliem2006}, the attention of the solar physics community was drawn to a different type of `lateral kink' instability, the so-called torus instability~\cite{Shafranov1966,Bateman1978,Chen1996}. This torus instability corresponds to the tendency of a toroidal ring current to expand uniformly due to its (Lorentz) hoop force, and is unavoidable when the external (poloidal) magnetic field variation -- which normally opposes this expansion -- decreases fast enough. The analysis in~\cite{Kliem2006} used a large aspect ratio (hence $\epsilon\ll 1$) whole or partial current loop, neglected gravity, pressure and external toroidal (i.e. along the current loop) magnetic fields, and ignored all aspects of the internal variation of equilibrium quantities within its minor radius $a$. Using an external poloidal field variation as $B_{ex}(R)\propto R^{-n}$ for the region beyond the current loop $R>R_0$, the critical decay index $n_{cr}$ turns out of order $3/2$, making $B_{ex}(R)$ variations where $n>n_{cr}$ unstable to the torus expansion. It was argued convincingly how this criterion unifies findings applicable to spheromak type laboratory findings, as well as for line-tied solar coronal loops that erupt and lead to Coronal Mass Ejections (CMEs). The line-tying is heuristically accounted for when considering only a partial loop section, and as soon as the loop shape is at least semicircular, line-tying is expected to enhance its expansion. For the CME relevant case, the decay index is to be quantified from the detailed variation with height of those field lines overlying a flux rope. In reality, such overlying, restraining field variation may not be purely poloidal, and for the solar corona one usually relies on photospheric magnetic field extrapolations to quantify its topology. In~\cite{Demoulin2010}, a clear link was found between the torus instability point-of-view, and a frequently invoked current-circuit inspired model, where a current channel of increasing intensity eventually loses equilibrium at a given height. This allowed one to link results for both circular shaped and straight current channels, and identified critical indices in the range [1.1,1.3] for current channels of observed or simulated thickness.

\section{The usual suspects: observational and numerical confirmations}

\subsection{Kink instability}
\begin{figure*}
\centerline{\includegraphics[width=\textwidth]{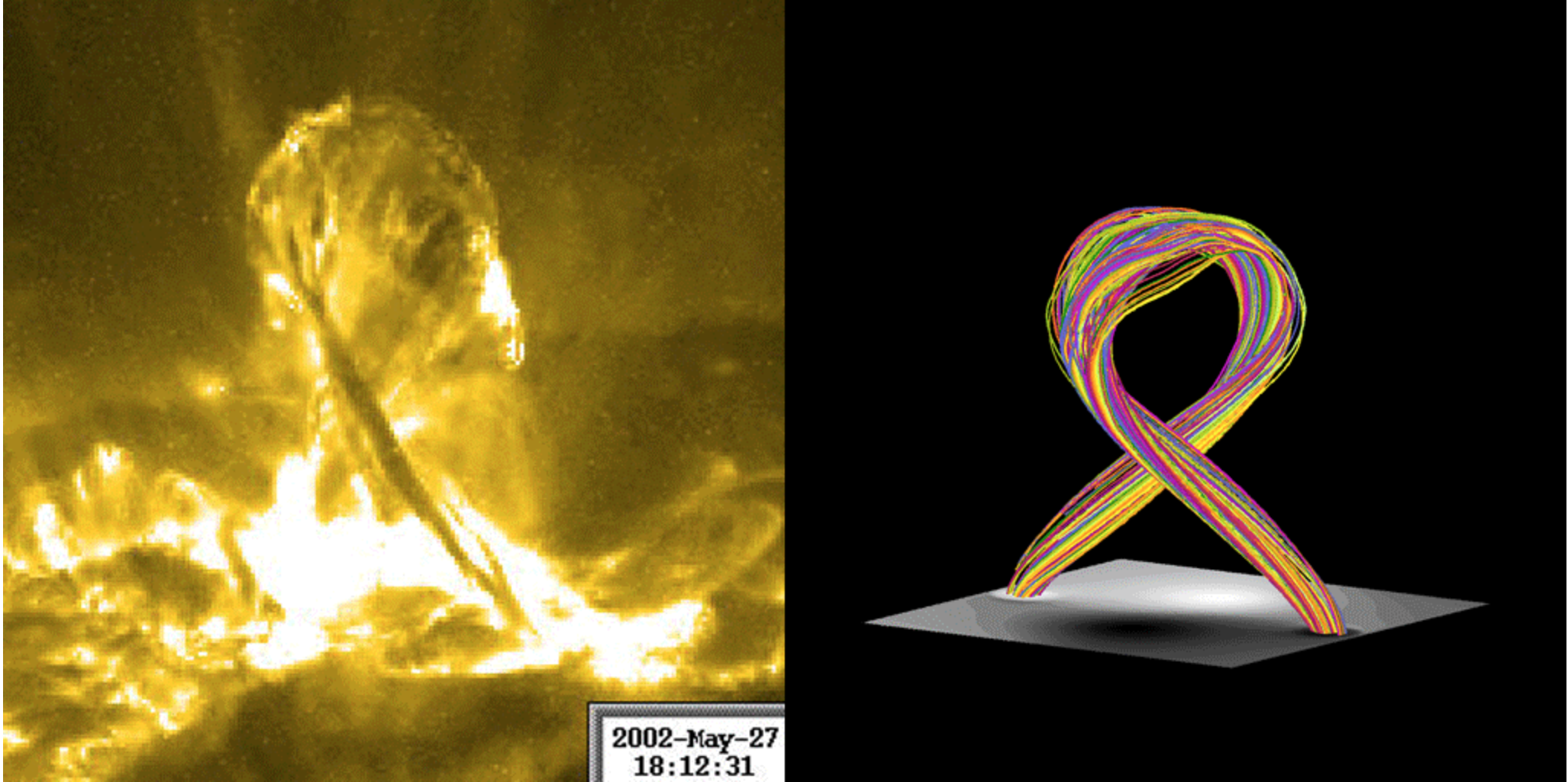}}
\caption{A TRACE observation (left) of a confined eruption, resulting from a kink unstable flux rope. The right panel shows the magnetic field deformation due to the kink, simulated in zero-beta approximation, from~\cite{Torok2005}.}\label{f-KI}
\end{figure*}

Perhaps the most striking evidence for kink instability underlying solar activity was provided by~\cite{Torok2005}, who analysed and reproduced both a confined eruption (i.e. failed CME) and a fast CME event. Using the analytical 
(partial) torus-shaped flux rope model from Titov-D\'emoulin~\cite{TD1999} as initial condition for a zero-$\beta$ (and no gravity) numerical MHD experiment, both the confined eruption case (see Fig.~\ref{f-KI}) and the ejective case were recovered in their overall morphological properties during the flux rope rise. Figure~\ref{f-KI} contrasts a TRACE 195 $\AA$ image with the simulated field topology when the failed eruption is at its peak: the helical deformation due to the kink onset is evident. Since then, many observational~\cite[e.g.]{Liu2016,Srivastava2010} and numerical findings~\cite[e.g.]{Hassanin2016,Amari2018} further corroborated the role of kink instabilities for initiating single flux rope eruptions. E.g., \cite{Srivastava2010} used multi-wavelength observations to estimate an unusually large twist value $\Phi\approx 12 \pi$ in an active region coronal loop, giving rise to flaring as related to the then unavoidable kink evolution. 

The kink instability induces the formation of a helical current sheet wrapping around the flux rope. A further parametric study of the role of kink instability for a confined eruption in AR 9957 was presented in~\cite{Hassanin2016}, and showed how the reconnection between the helical current sheet and the overlying field may destroy the flux rope while rising. A second phase of reconnection between the original flux rope footpoints then leads to a reformation phase, potentially allowing homologous eruptions. Although the model from~\cite{Hassanin2016} exploits zero-beta conditions and can only identify (numerical) reconnection sites of interest for true flaring processes, they do relate well to observed sites of brightening at the boundary of erupting flux rope footpoints~\cite{Wang2017} or cross-sections~\cite{Gou2019}.

A major point to note from the theoretical stability considerations is that internal equilibrium variation and also pressure effects (i.e., finite-$\beta$) can significantly influence stability thresholds. To study the role of internal pressure variation on kink stability for single flux rope structures, a parametric survey was conducted by~\cite{Mei2018} where a modified Titov-D\'emoulin~\cite{TD1999} configuration was adopted, such that one could vary pressure contributions, aspect ratios, and thereby modify the expected liability to either internal versus external kink disruptions. In 3D isothermal MHD simulations, the internal plasma pressure adds a tire tube force~\cite[chapter 4, page 77]{Freidberg87} (aiding the expansion already invoked in the torus instability, while the original analysis~\cite{Kliem2006} exclusively resorts to the Lorentz hoop force), and a toroidal field component in the flux rope also adds curvature that leads to a downward force on the flux rope. One can construct magnetic flux ropes that use these opposing forces to achieve internal flux rope equilibrium, which thus have a specific plasma-$\beta(r)$ and $q(r)$ profile. Line-tying was enforced at the lower boundary, and scenarios evolving from external kink, or from combinations of internal and external kink instabilities were demonstrated. In the latter, mixed scenarios, reconnection sites are found both within and external to the flux rope system, and sigmoidal current structures appear naturally, without invoking flux emergence. 

\begin{figure*}
\centerline{\includegraphics[width=\textwidth,height=10cm]{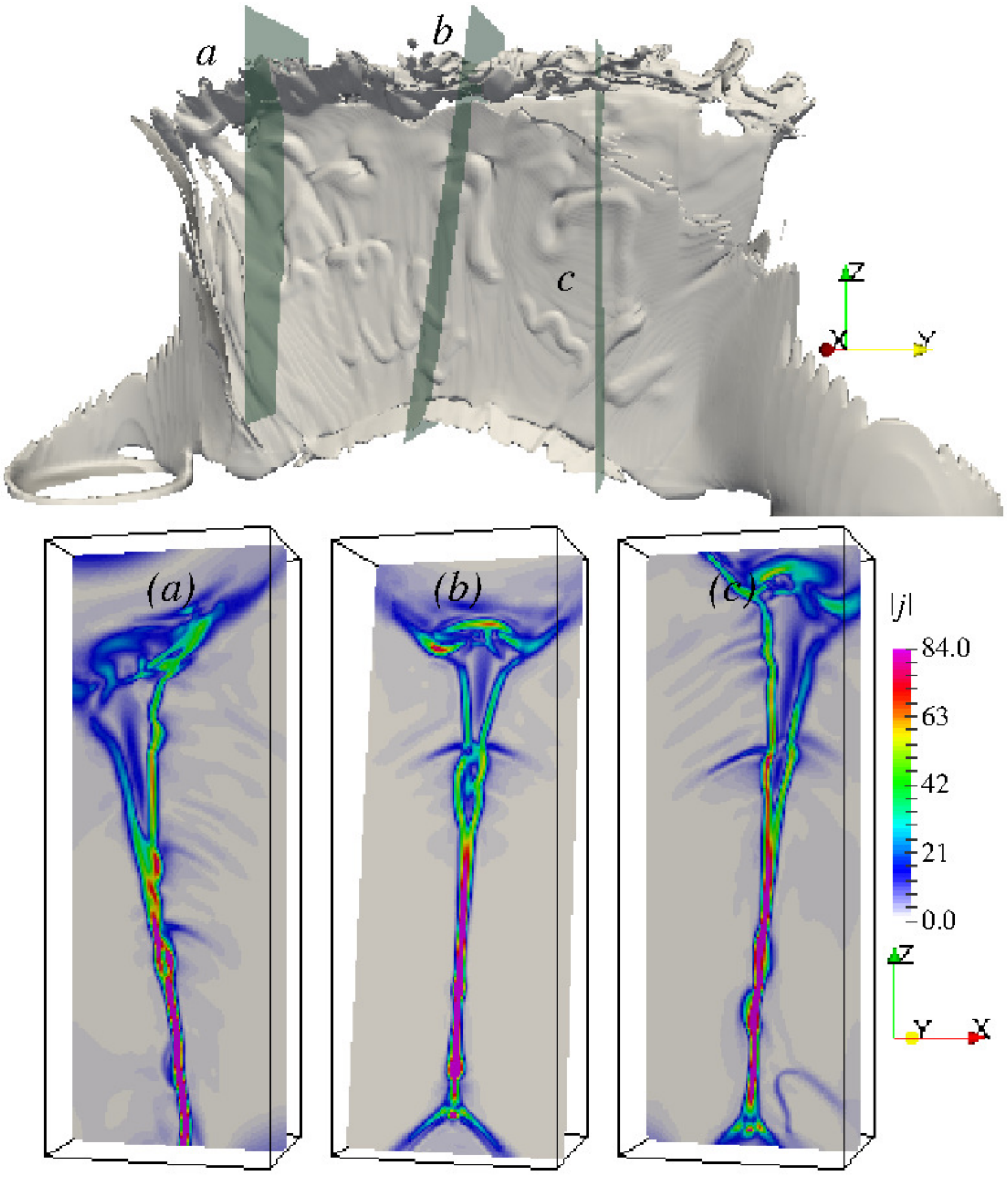}}
\caption{The current sheet details, obtained by a finite beta, isothermal MHD simulation from~\cite{Mei2017}, underneath the kink-unstable flux rope evolution.}\label{f-Mei}
\end{figure*}

Switching from zero-$\beta$ to finite-$\beta$ conditions is not merely a technical issue, but in fact is a necessity to address the reconnection layer physics in more detail. Indeed, all zero-$\beta$ models necessarily miss out on reconnection aspects related to slow mode shocks: the zero-$\beta$ assumption eliminates all slow waves from consideration. Using an isothermal, modified Titov-D\'emoulin~\cite{TD1999} model, whose twist ensures external kink instability \`a la Kruskal-Shafranov, the technique of adaptive mesh refinement was exploited by~\cite{Mei2017} to simulate the reconnection layer underneath the erupting flux rope in unprecedented detail. Figure~\ref{f-Mei} shows the structure of the current sheet at a representative instant, along with several 2D cross-sectional views on the current distribution. The finite-$\beta$ conditions allow to capture the 3D generalization of the Petschek-type slow shock configurations~\cite{Petschek1964}, as well as the island structures already shown in the 2D simulation from Fig.~\ref{f-FR}.
These islands now are seen to extend as finite length flux rope substructures within the current sheet, and arise from the tearing type instability that triggers chaotic island dynamics. The simulation relies purely on numerical resistivity (i.e. deviations from flux-freezing entirely due to discretization errors) to achieve reconnection, but could estimate that the adaptive grid roughly achieved an effective Lundquist number of $10^4$ near the current sheet. This estimated Lundquist number quantifies resistivity through $v_A L/\eta_{\rm{num}}$, where $v_A$ indicates the typical (e.g. footpoint) Alfv\'en speed and $L$ the flux rope length. The numerical resistivity $\eta_{\rm{num}}$ was taken from the obtained reconnection inflow speed and current sheet width. In general, the Lundquist number relates directly to the magnetic Reynolds number where $R_m=vL/\eta$, where $v$ is a characteristic speed, e.g. the sound speed.

\begin{figure*}
\centerline{\includegraphics[width=\textwidth]{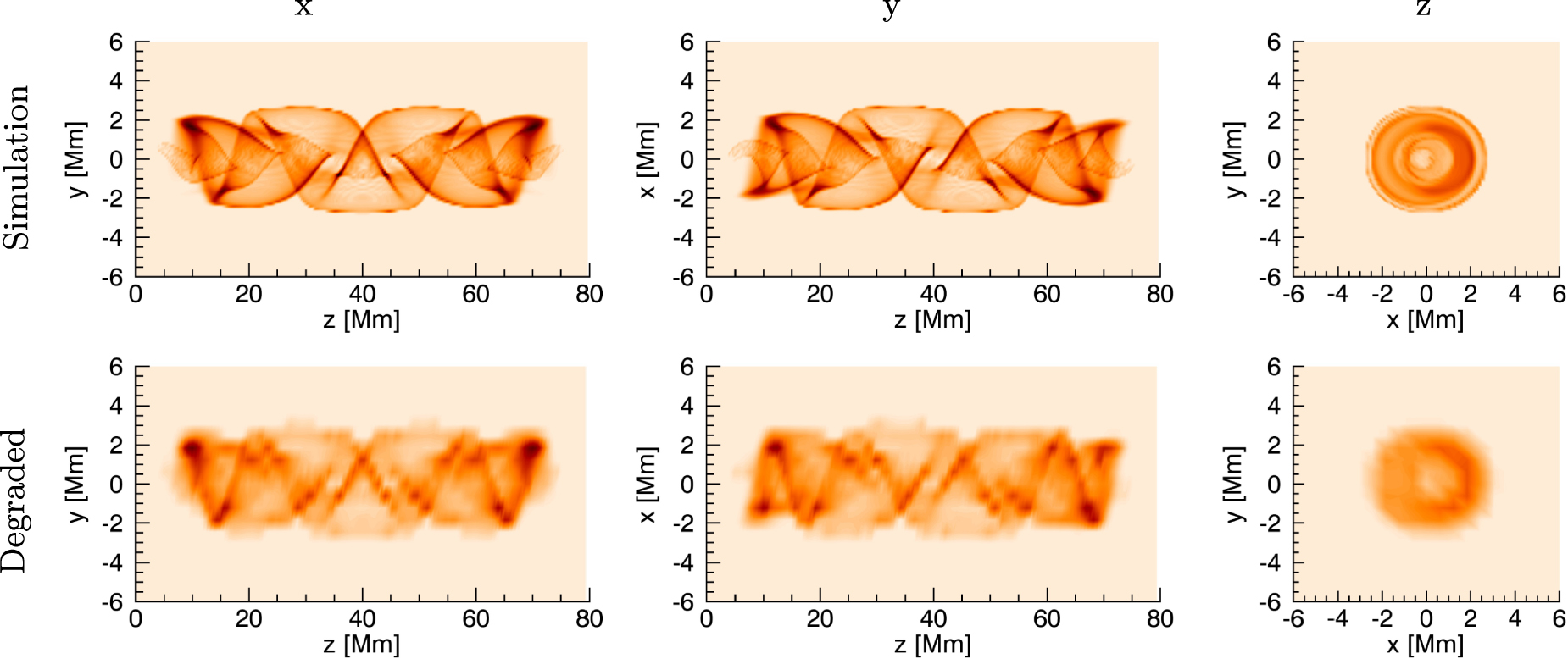}}
\caption{Synthetic views on an MHD simulation of a kink unstable cylindrical flux tube~\cite{Snowetal2017}. The tube is line-tied at both ends (along $z$), and the intensity of the Fe X line for Hinode/EIS observations is synthesized. The bottom panels are degraded to 1 arcsecond resolution.}\label{f-Snow}
\end{figure*}

Another aspect which requires at least finite-$\beta$ MHD simulations, preferably with all relevant energetics included, is forward modeling: virtual observations mimicking specific instruments on modern facilities. Using a 3D resistive MHD simulation of a highly kink-unstable, line-tied cylindrical flux rope~\cite{Bothaetal2011}, incorporating field-aligned thermal conduction (but no gravity), a serious attempt to synthesize intensity and Doppler maps for Hinode/EIS observations was performed~\cite{Snowetal2017}. The setup was inspired by the highly twisted coronal loop evolution mentioned earlier~\cite{Srivastava2010}, and the Doppler maps do reveal the velocity patterns induced by the kink development. Also the loop expansion in the radial direction, and an increase of the spectral line intensities near the loop edge, is consistent with the kink deformation. A representative intensity map for the Fe X spectral line (in three different integrated views along $x$, $y$ and $z$ respectively) is shown in Fig.~\ref{f-Snow}. The synthetic view on the simulation is also degraded to 1 arcsecond resolution, where obviously much fine-structure is no longer detectable (particularly true for localized reconnection events).

\subsection{Torus instability}
\begin{figure*}
\centerline{\includegraphics[width=0.6\textwidth]{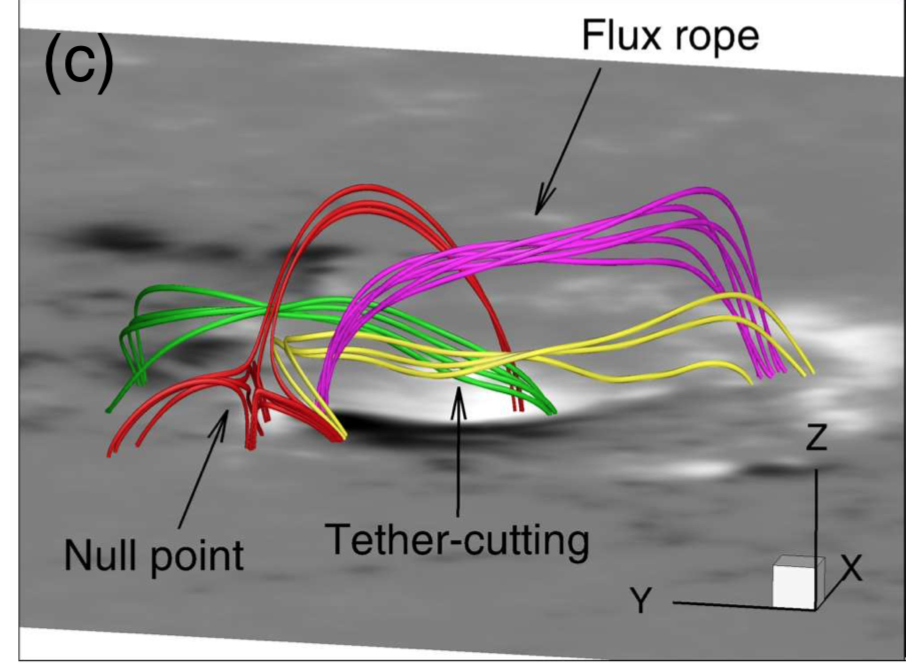}}
\caption{A nonlinear force-free field extrapolation for the flaring region analysed by~\cite{Zou2019}: the purple flux rope only erupts much later, despite successive reconnections in its vicinity. This is consistent with a failure to exceed the torus instability treshold.}\label{f-Zou}
\end{figure*}

The torus instability criterion has also become a standard diagnostic tool to test whether simulated~\cite[e.g.]{Fan2010} or observed~\cite[e.g.]{Liu2008} eruptions follow from an overlying magnetic configuration that is too weak to withhold expansion. 
In fact, observational evidence~\cite{Guoetal2010} supports interpretations where kink instabilities merely act as initial driver of solar eruptions, whereby the difference between confined and eruptive scenarios ultimately comes from torus-unstable configurations: torus instability seems a requirement to explain actual eruptions \cite{BaumgartnerThalmann2018,WangLiu2017}. \cite{WangLiu2017} quantified decay indices above polarity inversion lines in 60 flare events, and found that the critical height (where the index $n$ goes above 1.5) is significantly higher for confined flares than for eruptive ones. A similar statistical significance was found when analyzing 44 class M5-and-above flares~\cite{BaumgartnerThalmann2018}. Meanwhile, actual eruptive events have also been linked conclusively with the torus instability criterion. In~\cite{Zou2019}, a detailed analysis of an X-class flaring region was performed (NOAA 12673 in September 2017). An X2.2 flare showed clear evidence for a two-step reconnection process with corresponding flare brightenings, one involving a null-point reconnection and the other due to a tether-cutting reconnection. A nonlinear force-free field extrapolation was performed to get quantitative insight into the overall active region field topology, thereby identifying an expanding magnetic flux rope (purple in Fig.~\ref{f-Zou}). Despite the two-step reconnection occurring in its vicinity, the flux rope failed to erupt, and this was found to be consistent with the torus instability threshold: strapping flux prevented the eruption. Later on, the same active region showed an eruptive X9.3 flare, and then the instability threshold got exceeded.

\begin{figure*}
\centerline{\includegraphics[width=\textwidth]{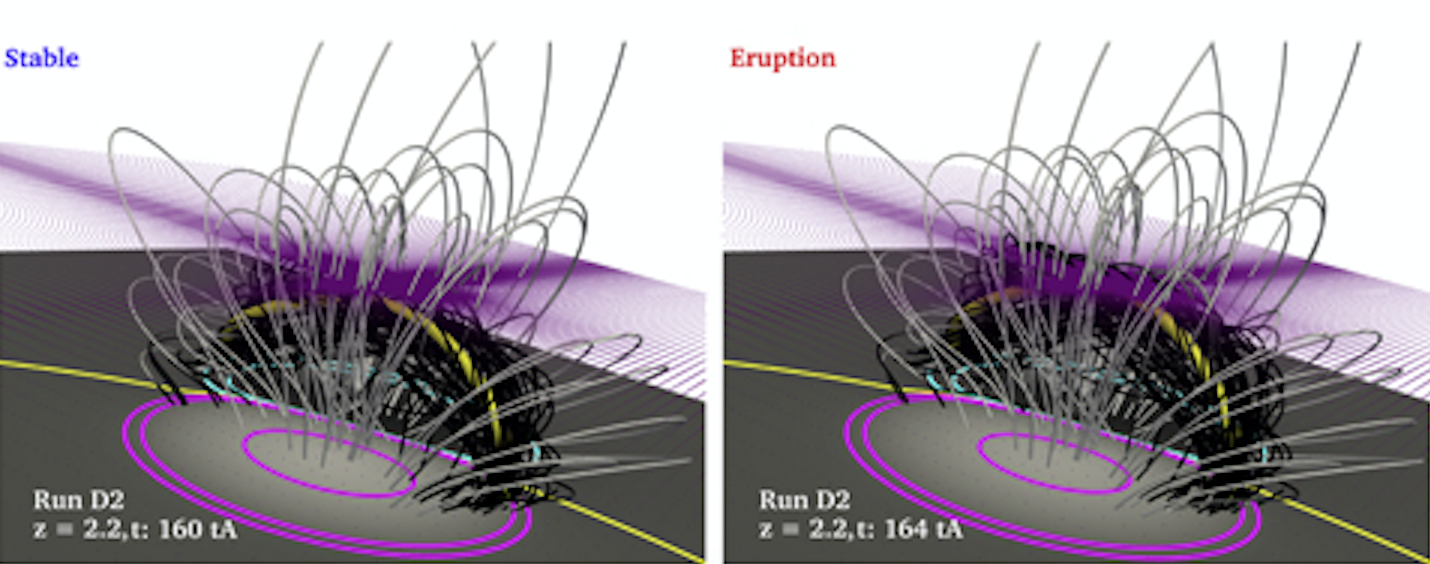}}
\caption{Verifying the torus instability threshold, with zero-$\beta$ simulations from~\cite{Zuccarello2015}. The left case is stable, as the entire flux rope (with black field lines, the yellow field line indicating its axis) is below the critical index height, indicated with the purple cutting plane, while the right case leads to eruption.}\label{f-Z}
\end{figure*}

As stated earlier, the critical stability threshold analysis can only be carried out for a highly simplified wire current setup~\cite{Kliem2006}, where internal flux rope variations are ignored, and the threshold relates to the variation of the overlying magnetic field with height. This can be improved upon with zero-$\beta$ MHD models, where the current ring is replaced by a true flux rope with internal magnetic field variation. In~\cite{Zuccarello2015}, this was done in a series of numerical experiments, which deformed initial potential fields to current-carrying, erupting flux ropes by shearing and converging motions on the bottom boundary. At all instants in the MHD evolution, one can identify the flux rope, once formed, and directly compute the local decay index of the overlying field variation. Figure~\ref{f-Z} shows the field configuration for a stable (left) and unstable, i.e. erupting (right), realisation. The flux rope is shown together with a (purple) critical height surface, indicating the location where the critical decay index of 1.5 is reached.  The left case does not erupt when the driving motions stop, which is consistent with the entire flux rope being below the critical height surface. The case at right partly did extend above the critical height surface and indeed erupts fully, even though the flux rope axis (yellow in Fig.~\ref{f-Z}) did not yet reach the critical height. 

Another recent example in favour of torus instability was provided by a data-driven zero-$\beta$ simulation in~\cite{Guo2019}. There, the MHD evolution of active region 11123 from November 2010 was reproduced numerically, using both vector magnetic field HMI data and a derived vector velocity field in a spatio-temporally evolving bottom boundary. The model provided insight in the various phases of the flux rope eruption: a gradual rise resulting from magnetic reconfigurations in its vicinity preceded a changeover where the flux rope starts to rise at several tens of ${\rm km/s}$.  Analysing the instantaneous decay index above the rising flux rope, the changeover from slow to fast rise and ejection was found consistent with the critical index crossing. 

As a final note on the torus instability, a recent laboratory experiment~\cite{Myers2015} revealed the role of ambient toroidal fields, which were ignored in the original analysis. It was found that such fields could prevent the flux rope eruption, and hence lead to `failed torus' events, although the usual torus-instability criterion would be met. The orientation of the overlying field is also known to be important on theoretical grounds, as pointed out in~\cite{Isenberg2007}, where a Titov-D\'emoulin (TD) setup was investigated with respect to stability against line-tied perturbations. A detailed force analysis showed that the flux rope can rotate during eruption, out of the original plane of the toroidal loop, associated with the details of the submerged line current used in the TD model. The details of both internal and external field variations, as well as flow and all thermodynamic conditions, are thus very important for deciding on the ultimate fate of confined versus eruptive CMEs, and the stability criteria that usually derive from idealized setups, must always be scrutinized. A recent observational counterpart to the laboratory finding on failed torus events was presented in~\cite{Jing2018}, where 38 flare events (stronger than M5) were categorized in the corresponding decay-index $n$ versus twist number $T_w$ parameter space. The decay index relates to torus instability (quantifying the potential strapping field above flux ropes), while the twist number is kink relevant and derives from nonlinear force-free field extrapolations. The twist number could not differentiate eruptive from confined events, but the ejective cases consistently showed $n\gtrsim 0.8$. Another study by~\cite{Zhou2019} investigated 16 failed filament eruptions in detail, and hinted that significant rotational motion seems correlated with failed eruptions, that would normally be judged as torus unstable. Hence, even our current understanding of single flux rope events, and what factors differentiate between failed and true CMEs, is still heavily researched. In what follows, we turn attention to the more complex scenarios involving multiple flux ropes.

\section{Interacting current channels}
All modern EUV or X-ray views on the active solar corona show that the theoretical isolated flux rope system is a poor representation of reality: myriads of loops coexist at the same time, and they likely correspond to multiple current-carrying flux ropes embedded within more potential field regions. In fact, whenever a tether-cutting reconnection scenario is invoked~\cite{Moore1980}, a reconnection below an existing flux rope takes place such that overlying fields that `tether' it to the photosphere are cut and allow an escape. When drawing this in a 3D scenario, it typically reconfigures a quadrupolar field $\oplus_1-\ominus_1-\oplus_2-\ominus_2$ situation where two neighbouring flux ropes have their opposing polarity feet (i.e. $\ominus_1$ and $\oplus_2$) near one another. The tether-cutting changes the original connectivity from $\oplus_1\leftrightarrow\ominus_1$ and $\oplus_2\leftrightarrow\ominus_2$ to $\oplus_1\leftrightarrow\ominus_2$ and $\ominus_1\leftrightarrow\oplus_2$. In practice, this involves two nearby current-carrying flux ropes to interact and reconfigure. Such a scenario was e.g. invoked in the work by~\cite{Zou2019}, where tether-cutting between the green and yellow flux systems underneath the (purple) flux rope was identified, as shown in Fig.~\ref{f-Zou}. The topological reasoning based on nonlinear force-free field extrapolations misses out on more dynamical MHD aspects, that have started to incorporate interacting current channel aspects. Such aspects are discussed next.

\subsection{Slingshot reconnection}
In~\cite{Linton2001}, a numerical investigation of colliding and reconnecting flux ropes was presented, where the type of interaction that resulted related to the angle between both rope axes, and whether they carried the same or opposite helicity (i.e. current orientation). Flux tubes that collided were either bouncing, merging, tunneling, or showed what was introduced as `slingshot reconnection'.  A set of identical force-free flux tubes were exploited in 3D compressible MHD simulations, chosen to be stable to current-driven kink modes. The tubes were forced to interact by imposing a converging stagnation point flow, and a triple periodic box setup ignored any effects of line-tying. If the flux tubes were nearly parallel, and had antiparallel currents, a bounce interaction resulted. For parallel flux tubes with parallel currents, a merging of the two tubes into a single widened flux tube was found. These findings are consistent with the fact that the bounce scenario brings locally parallel (azimuthal) magnetic field lines into contact, while the merge scenario meets up antiparallel (azimuthal) field, provoking reconnection. When two flux tubes meet up under an angle, with both their azimuthal and axial field being antiparallel, the field topology lends itself to even more pronounced reconnection, where the flux ropes can exchange their connectivity completely: a slingshot reconnection results. The `slingshot' analogy is made since the tension force of the reconfigured flux systems is expected to expel the reconnected ropes away from the interaction zone.

\begin{figure*}
\centerline{\includegraphics[width=0.49\textwidth]{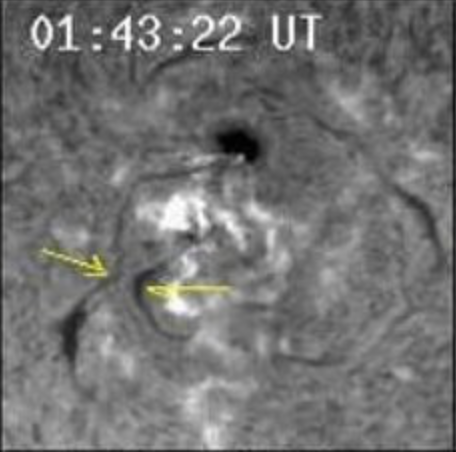}\includegraphics[width=0.49\textwidth]{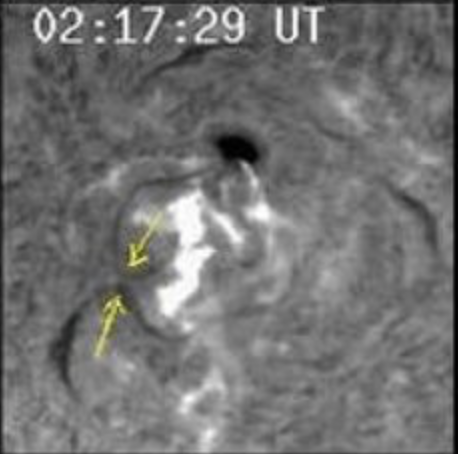}}
\caption{H-$\alpha$ views on an interacting filament system, from~\cite{Kumar2010}. Of particular interest is the double filament system indicated by the yellow arrows: it is clearly seen how the filaments change their connectivity from the left to right panel, indicative of slingshot reconnection.}\label{f-obs}
\end{figure*}

A particularly interesting observation of two interacting filaments provided clear evidence for slingshot reconnection, as the filaments seen in H-$\alpha$ merged at their middle section, to disconnect in a seemingly changed topology afterwards~\cite{Kumar2010}. Figure~\ref{f-obs} shows the filament configuration  before and after this reconnection. Numerical evidence for the slingshot scenario at work was presented in~\cite{Torok2011}, using two adjacent (line-tied, kink-stable) Titov-D\'emoulin~\cite{TD1999} flux ropes with antiparallel currents along their axes. The zero-$\beta$ simulation employed various stages of (i) relaxation, (ii) diverging flows to weaken the flux systems between the two ropes, and (iii) a converging flow to bring the middle flux rope sections closer to each other. In the latter phase, the middle parts of both ropes reconnected in accord with the slingshot reconnection process, whereby the foot point connections of the two ropes were mutually exchanged. The simulation provided favourable evidence for the filament evolution, in the sense that the dipped field line regions in the MHD evolution behaved similar to the observed H-$\alpha$ reconfiguration. As several important (e.g. all thermodynamic and stratification) aspects were ignored, the details of reconnection could not be discussed, as they arose from numerical diffusion. 

\subsection{Further clues from interacting filaments}

The slingshot reconnection identified observationally in~\cite{Kumar2010} may well be a relatively rare event, since a follow-up simulation study~\cite{Linton2006} highlighted that this slingshot process is more complex when non-identical flux ropes interact. For tubes with unequal flux, the merge process between parallel current systems behaved similar to the idealized, identical flux setup. However, the slingshot scenario can lead to rather different reconnection behavior: the smaller flux tube may only interact with the outer shell of the larger flux tube. Observations in support of merging reconnection can be found in~\cite{Jiang2014}, where two sinistral filaments merged into a single sinistral filament. Interacting, so-called double-decker filaments~\cite{Liu2012}, demonstrate intricate mass and flux transfer events, which may also happen between chromospheric fibrils and single prominences, as found observationally by~\cite{Zhang2014}. Mass and magnetic flux transfer between overlying branches of a solar filament were found to induce equilibrium loss of the upper filament branch in~\cite{Zhu2014}. Another double-decker filament configuration showed a clear coalescing of both filaments~\cite{Zhu2015}, as the lower one erupted while the upper one descended. The various ways in which filament-filament interactions are observed to proceed seem to involve intriguing MHD aspects of interacting, separating and/or coalescing flux tubes with mass and flux transfer events, mitigated by (partial) reconnection.

\subsection{Sympathetic eruptions}

\begin{figure*}
\centerline{\includegraphics[width=\textwidth,height=10cm]{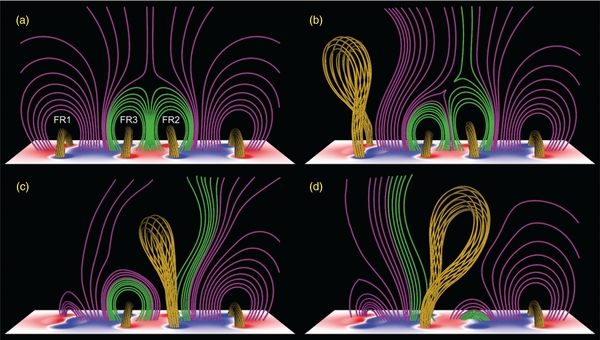}}
\caption{Sympathetic eruptions in multiple flux rope systems, from~\cite{Torok2011b}. The bottom boundary quantifies the multipolar magnetic field, where initially four flux ropes are found with overlying field variations as indicated by the purple and green field lines. The leftmost flux rope is gradually brought to torus instability, and its eruption triggers sympathetic eruptions of the neighbouring flux ropes.}\label{f-TT}
\end{figure*}

If adjacent flux ropes are far apart, eruption of one magnetic flux rope due to e.g. kink instability, can modify the overlying flux for neighbouring flux ropes found in streamer topologies. This then may drive a sympathetic eruption of the flux ropes within the streamer. Figure~\ref{f-TT} shows snapshots of a zero-$\beta$ simulation~\cite{Torok2011b}, which actually initializes 4 neighbouring flux ropes in an almost left-right symmetric setup. All four are stable to the kink and torus instabilities, but a controlled eruption due to torus instability is ignited on the leftmost flux rope by means of converging flows that expand the restraining overlying field. As this erupting flux rope interacts with the overlying field of both flux ropes within the streamer, these in turn become liable to the torus instability. This scenario has been invoked by~\cite{Wang2018}, where both a failed and a successful filament eruption occurred within a short interval, and where the observational indications for reconnection indicate a strengthening and weakening of the overlying field, respectively. 

\subsection{A new kid on the block: tilt and tilt-kink evolutions}

\begin{figure*}
\centerline{\includegraphics[width=0.32\textwidth]{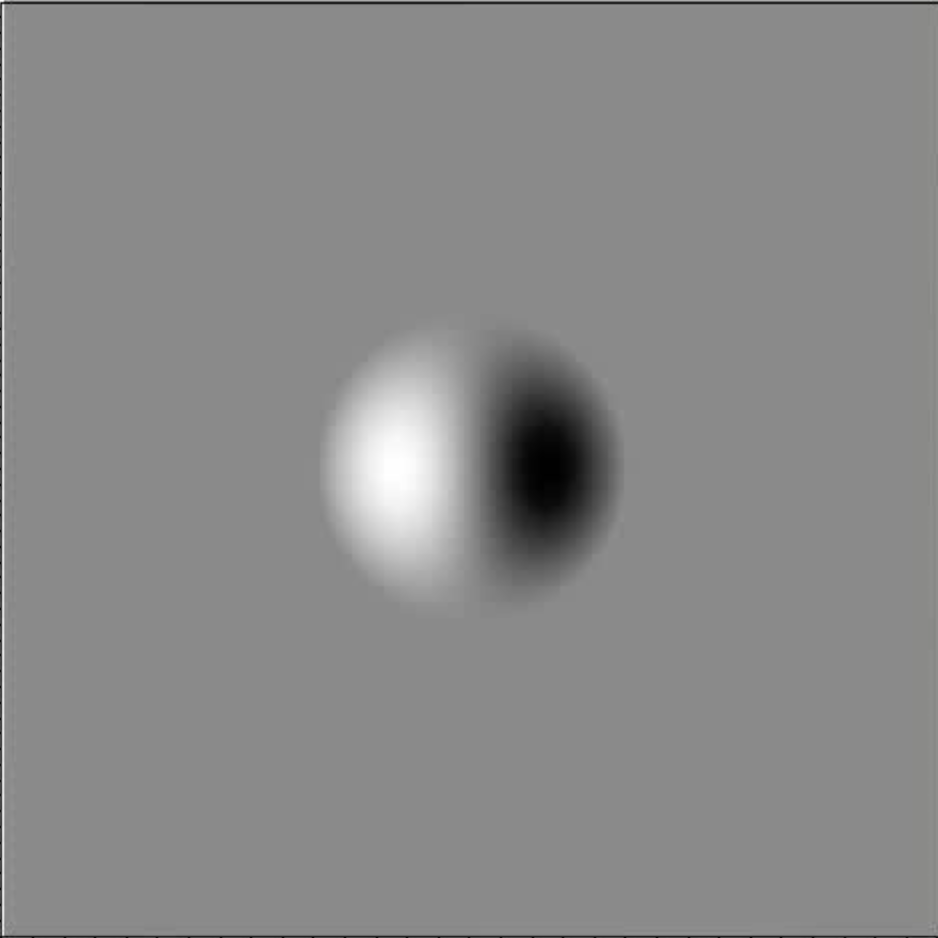}
\includegraphics[width=0.32\textwidth]{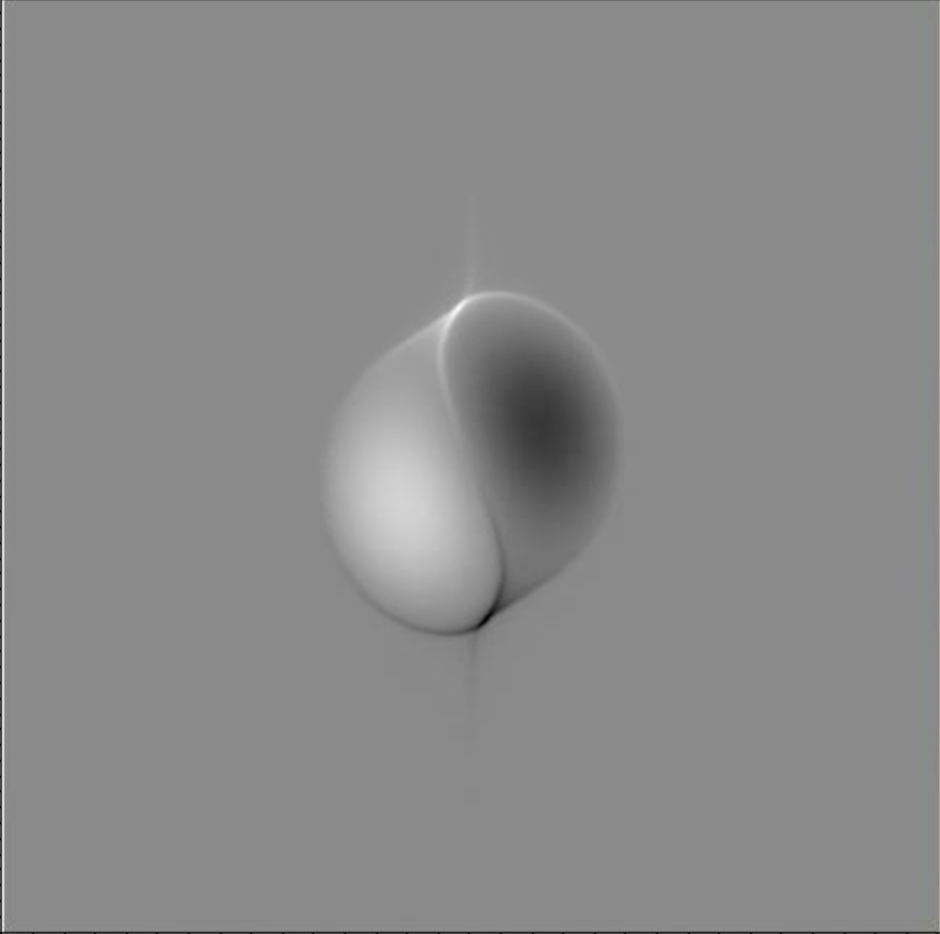}
\includegraphics[width=0.32\textwidth]{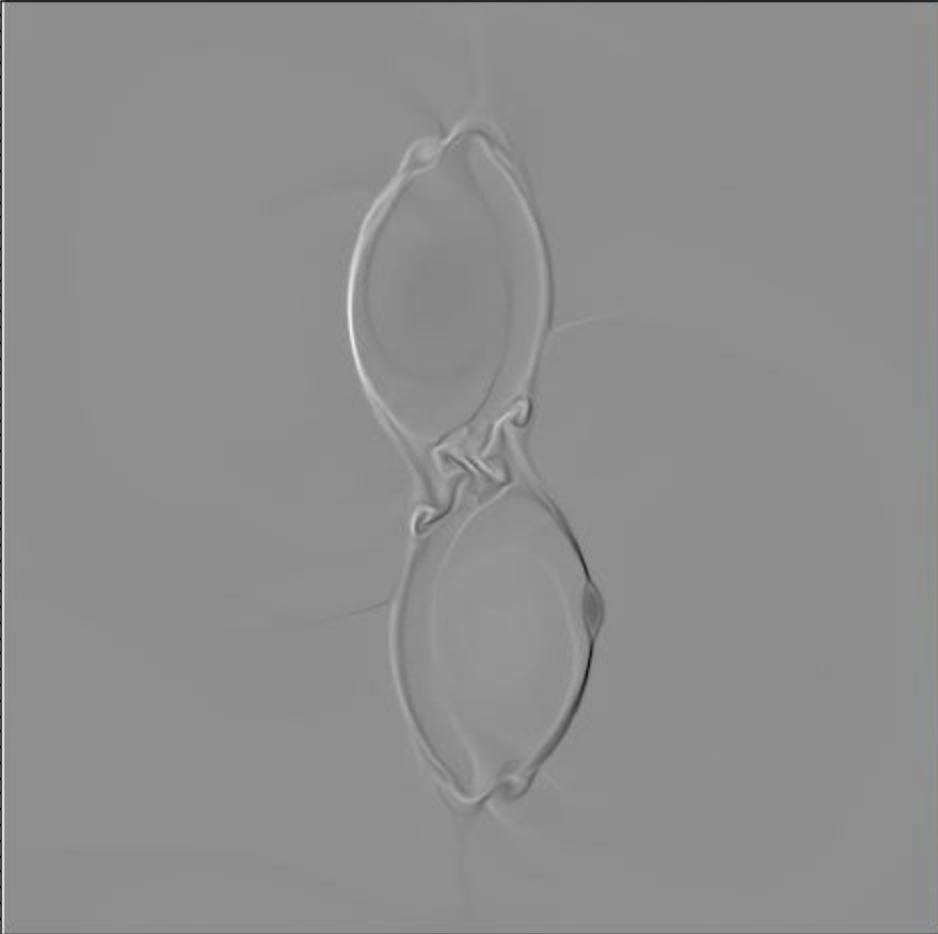}}
\caption{The basic out-of-plane current evolution in a tilt unstable scenario of two adjacent magnetic islands. The islands repel and rotate away from each other. The background field is uniform and vertical (along $y$) in this setup.}\label{f-R1}
\end{figure*}

Whenever interacting current systems are close enough, the analogy with parallel or antiparallel wire current systems instantly reminds us of Lorentz-force mediated interactions: parallel current wires attract, while antiparallel current wires repel. This basic physical fact seems largely ignored in current interpretations of coronal loop dynamics, but it is behind two MHD processes that are mirror images of each other: the coalescence of neighbouring magnetic islands or flux tubes when two currents run parallel~\cite{Finn1977}, and the tilt instability when the currents are opposing~\cite{Finn1981,Richard1990}.  In~\cite{Richard1990}, a rigorous analysis exploiting the energy principle confirmed that the tilt instability for adjacent antiparallel current islands operates on the ideal MHD timescale. This fact makes it an equally important macroscopic instability route, acting on the same Alfv\'enic timescale as the kink or torus instability discussed previously. Note in particular that the usual picture of two repelling currents does not fully apply: these would be expected to move directly away from each other, along the line connecting their centers of mass. Instead, the presence of walls (in the spheromak case) or a confining external magnetic field (as in the idealized setup from~\cite{Richard1990}) instead yields motions perpendicular to the line connecting the current centers: the islands then separate and rotate, see Fig.~\ref{f-R1}.

Observational evidence for the existence of nearby antiparallel current systems in the solar atmosphere is found in various works~\cite[e.g.]{Su2018,Liu2010,Awasthi2019}. In~\cite{Su2018}, consecutive flare events in NOAA active region 12396 were investigated, with 3D magnetic topologies reconstructed by two complementary non-linear force-free field methods. Loop systems with opposite helicity (i.e. antiparallel currents) were clearly identified in the complex evolution of this emerging active region. Furthermore, \cite{Liu2010} investigated gradually swelling coronal arcades, eventually leading to CMEs, where opposite helicity injection seemed a key ingredient during an extended period of quasi-static, pre-eruptive conditions. More recently, by analysing composite filament motions in NOAA 12685, \cite{Awasthi2019} deduced a double-decker magnetic hosting structure, with opposite helicity between a flux rope component and a sheared arcade. Such double-decker configurations have been reported first in~\cite{Liu2012}, where clearly superposed filament branches, seperated in height by 13 Mm, seemed consistent with either a double flux rope configuration, or a rope-arcade model. It was argued there that the eruption of this system is difficult to explain by invoking a helical kink instability, although it later did show a typical S-shape deformation. A zero-beta modeling effort~\cite{Kliem2014} rather supported the interpretation based on altered stability properties between neighboring flux ropes with the same current direction, involving flux transfer. We speculate here that an alternative interpretation could well be based on a tilt-kink evolution, invoking anti-parallel currents, as discussed in idealized model settings in what follows.

In recent work, the idealised planar 2D tilt-unstable setup from~\cite{Richard1990} has gradually been extended to full 3D~\cite{Keppens2014}, which is more relevant for actual coronal counterparts. The tilt instability is, similar to the kink and torus variants, a well-known phenomenon in laboratory plasma contexts: especially elongated spheromak fields are liable to tilt instability, where it represents a global rigid displacement of the plasma~\cite{Jardin1986}. In such spheromak setups, where we have an essentially axisymmetric configuration, a cross-sectional view containing the spheromak axis would show the plasma as two adjacent island structures. The instability affects the entire spheromak plasma shape, and appears as a relative displacement or `tilt' of these cross-sectional islands. The same instability persists in a configuration where two antiparallel current channels exist next to each other, so the axisymmetry of the spheromak can be replaced by two infinitely extended, parallel current-carrying cylinders. In that setup, it becomes representative for solar active region situations where two helical field bundles share the same overall orientation from a negative to a positive polarity magnetic field region, but have opposite helicities.

The purely planar, 2D setup from~\cite{Richard1990} is shown in Fig.~\ref{f-R1}, where the greyscale plots correspond to the out-of-plane current component (black and white indicating their antiparallel nature). The configuration starts from an exact ideal MHD equilibrium and the tilt instability proceeds to rotate and separate the two current systems. There is a background external constraining field in the vertical direction of these figures (note there is no gravity here): separating motions between the antiparallel current systems are constrained horizontally, but not vertically. Further in the nonlinear regime, strong localised, thinning current sheets develop along both repelling current channels, and tearing type reconnection events cause additional small-scale islands to appear. In between both separating current systems, turbulent interchange activity is seen. In~\cite{Keppens2014}, 2.5D realisations were investigated, which turn the purely planar $(x,y)$ case in a 3D setup where two helical flux ropes exist side by side. As long as the added axial $B_z$ component can not bend -- which is enforced when 2.5D assumptions (i.e. $\partial/\partial z=0$) are adopted -- there is little difference in growth rate for the tilt phenomenon, over a wide range of plasma-$\beta$: this was demonstrated for $\beta$ values ranging from order 10 down to order 0.1. These quoted beta values represent internal averages over both flux ropes, and their surrounding coronal beta values then become order 1 down to 0.05, approximately. What makes the tilt instability very relevant for flare related physics, is that it automatically generates singular current channels, that break up through tearing mode activity. This provides the needed conditions for particle acceleration, as will be discussed further on. The maximal current amplitude demonstrates a pronounced phase of exponential growth, and this makes the tilt evolution a challenging testcase for computational approaches, and ideally suited for grid-adaptive and/or order-adaptive simulations~\cite{Lankalapalli2007,Keppens2014}.

 \begin{figure*}
\centerline{\includegraphics[width=0.49\textwidth,angle=-90]{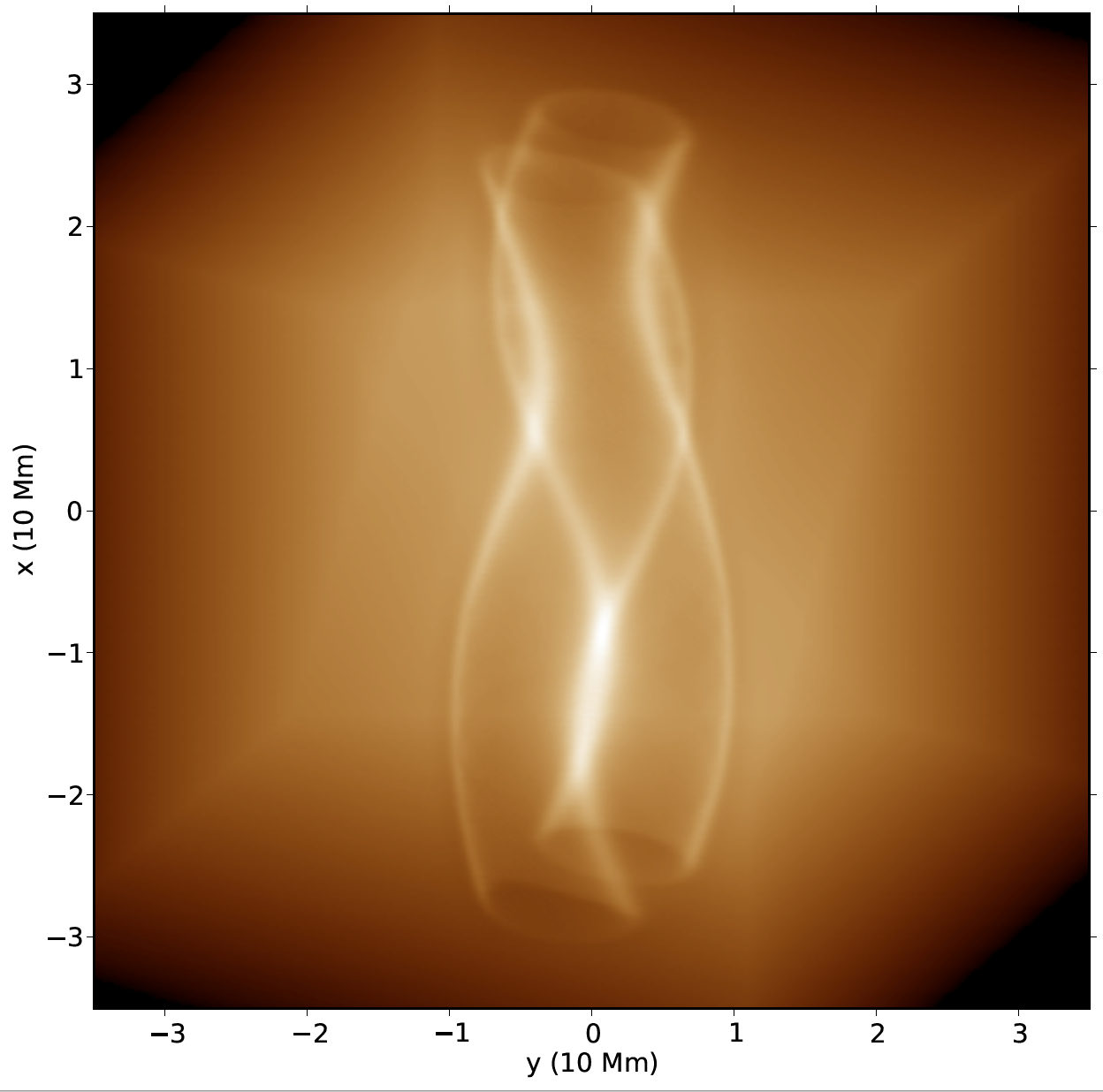}
\includegraphics[width=0.49\textwidth,angle=-90]{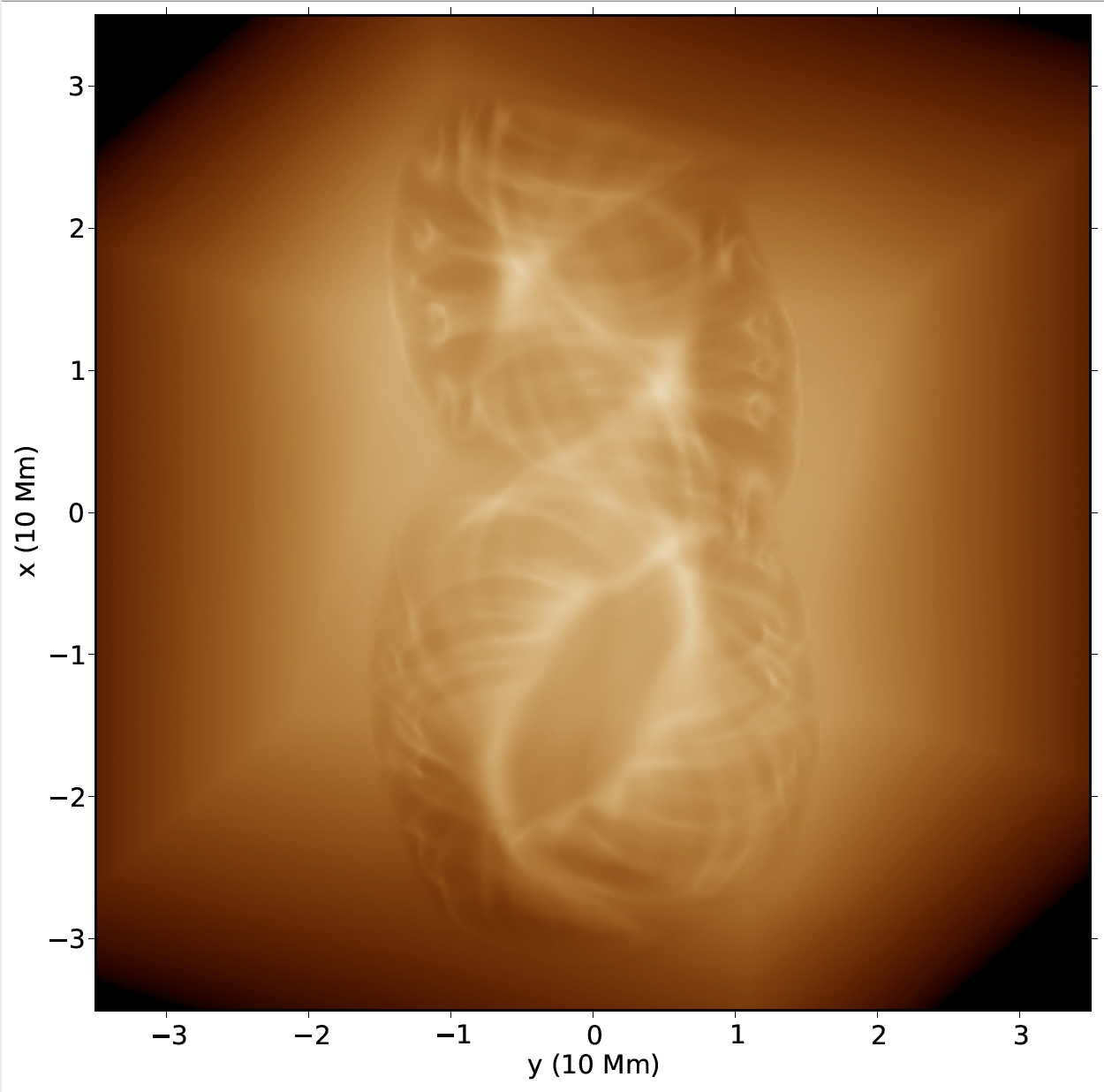}}
\caption{Tilt-kink evolutions in a system of two flux ropes with repelling currents. The resistive MHD simulation is translated to a synthetic EUV view, from~\cite{Keppens2014}.}\label{f-R2}
\end{figure*}

In full 3D, there is an interplay between the tendency for each twisted flux rope to be liable to kink instability, while the repelling nature of antiparallel currents drives tilt deformations. The liability to kinks still relates to the internal twist or $q(r)$ profile, and thus increasing $B_z$ will ultimately act as stabilizing for kinks. In~\cite{Keppens2014}, we found that indeed the kink deformation happens in both flux tubes, as long as the Kruskal-Shafranov condition is violated (no line-tying conditions were exploited), while one always observes the typical tilt scenario of rotation-separation. Taking the axial $B_z$ strong enough to render both tubes kink-stable would also prevent the tilt evolution to set in, although this conclusion may well be influenced by the combination of box size with wavelength of the imposed perturbation. A typical scenario, where the MHD result is translated to a virtual AIA 193 $\AA$ view is shown in Fig.~\ref{f-R2}. In the first stage, the two nearby current loops would be seen to swirl around each other, while separating (left panel). Later on, when also kink instability is possible, significant fine structure would be observed in each flux rope, due to the helical deformations of both ropes (right panel).

\section{Particle acceleration}

\subsection{Particle acceleration in kink unstable flux ropes}

The various MHD instabilities driving eruptions are accompanied by reconnection and heating events, which can potentially explain emissions in high energy (soft -- SXR -- and hard X-ray -- HXR) wavebands, including those from non-thermal particle populations. In~\cite{Pinto2016}, MHD simulations of kink-unstable flux ropes, in fully stratified, curved configurations were conducted, along with test particle computations.  Synthetic hard X-ray bremsstrahlung was synthesized, next to SXR thermal continuum views. It was pointed out that SXR emission features, usually loosely interpreted as outlining magnetic fields, misleadingly show different twist angles from the true magnetic field configuration. The study focused on confined, single loop evolutions, but incorporated all important non-adiabatic effects like anisotropic thermal conduction, realising a complete chromosphere-transition region-corona stratification. The HXR emission was found to be preferentially at the loop foot points, and the synthetic light curves for SXR and HXR emission recovered observed tendencies.  

\subsection{Particle acceleration in tilt-kink evolutions}

The tilt-kink scenario for solar flare particle acceleration aspects has also been investigated using the test particle approach, starting from 2.5D MHD simulations in~\cite{Ripperda2017a} and from 3D MHD scenarios in~\cite{Ripperda2017}. \cite{Ripperda2017a} used a variation of the two-island (i.e., two flux rope) equilibrium setup from~\cite{Keppens2014}, to ensure force-free conditions throughout each flux rope. With that modification, the magnetic field is purely poloidal external to both ropes, and rope-averaged plasma-$\beta$ values down to 0.04 could be explored. Exploiting adaptive mesh refinement, 2D simulations with effective resolutions up to $4800^2$ were feasible, and resistive effects due to a resistivity parameter $\eta$ of order $10^{-4}$ are fully resolved.

\begin{figure*}
\centerline{\includegraphics[width=0.49\textwidth,height=5cm]{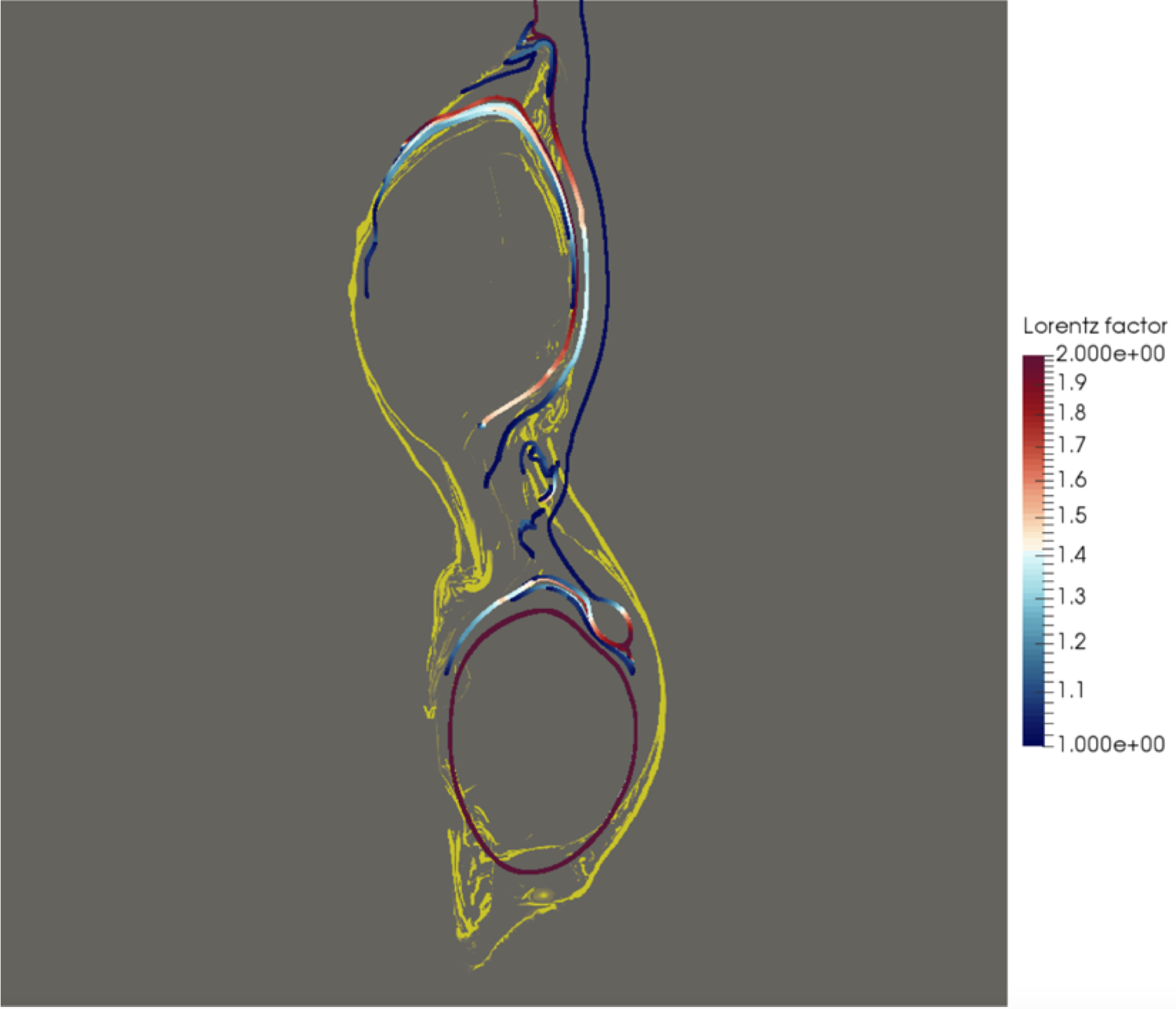}\includegraphics[width=0.49\textwidth,height=5cm]{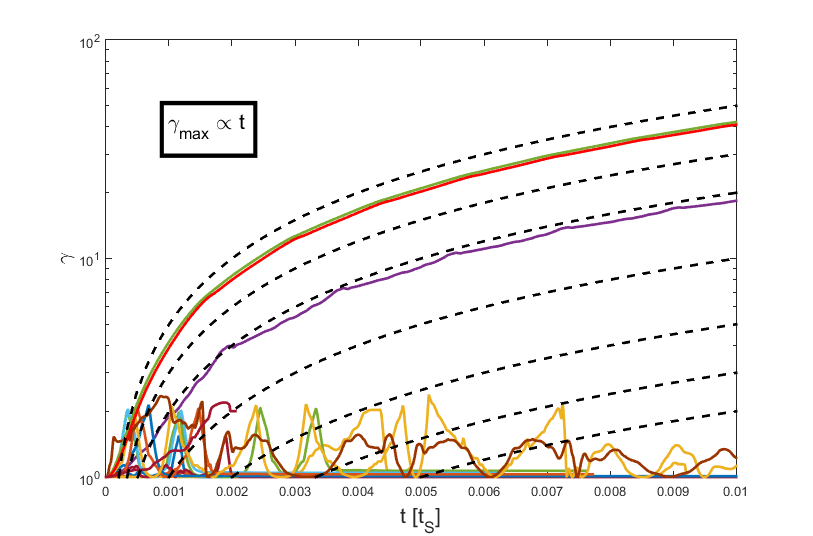}}
\caption{Tilt evolutions as particle acceleration engines: left panel shows the parallel electric field as background, along with selected proton trajectories. The latter are coloured by Lorentz factor. The right panel gives the time evolution of selected accelerated protons, showing two non-thermal populations: those accelerated indefinitely in the current channels, and those repeatedly encountering reconnection sites. From~\cite{Ripperda2017a}.}\label{f-B1}
\end{figure*}

In those detailed 2D MHD evolutions, charged particle populations of size 200000 were then subjected to the guiding centre approximation (GCA), employing a fully relativistic description. Figure~\ref{f-B1} shows an MHD snapshot at left, where representative proton trajectories are illustrated. The background colour scale shows the parallel electric field, responsible for (too) efficient magnetic field-aligned acceleration, even though this particular case used an anomalous resistivity prescription to suppress the artificial acceleration within both current channels due to the 2.5D assumptions (as found in similar studies of 2D reconnection evolutions by~\cite{Zhou2016}). The particle trajectories shown are for particles that occasionally encounter the local reconnection sites. At right, the evolution of the Lorentz factor for such non-thermal protons is illustrated. There, two classes of energized particles are seen: those whose acceleration is still unbounded by parallel electric fields (with $\gamma(t)\propto t$), and those showing fluctuating accelerations, repeatedly cycling through the reconnection regions.

\begin{figure*}
\centerline{\includegraphics[width=0.9\textwidth,height=5cm]{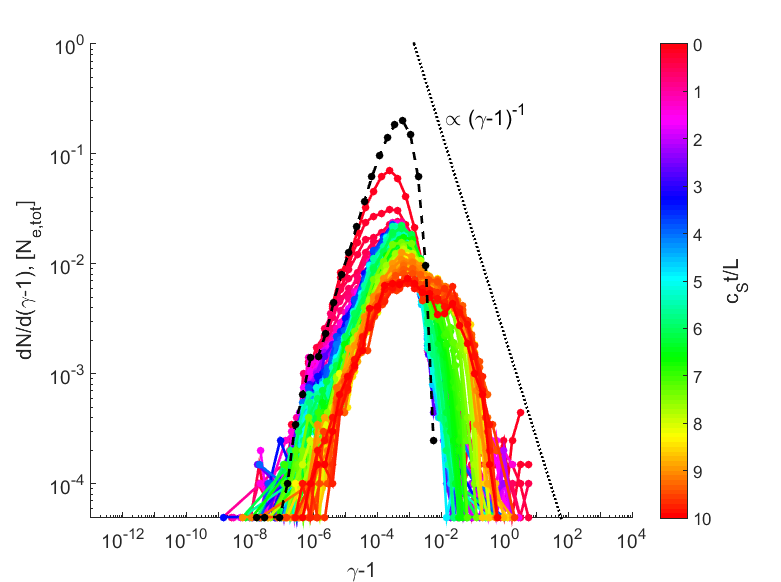}}
\caption{The temporal evolution of the energy distribution function for 20000 electrons, accelerated in a 3D tilt-kink scenario. A clear power law tail of non-thermal particles develops. From~\cite{Ripperda2017}.}\label{f-B2}
\end{figure*}

The same GCA approach was further adopted for full 3D tilt-kink evolutions~\cite{Ripperda2017}. To mitigate the indefinite acceleration when particles stay within the current channels, a thermal bath procedure was invoked: each time a particle cycles through the entire rope structure and encounters the periodic boundary, a thermal particle was injected at the opposing side. Careful analysis of particle trajectories identified distinct possibilities for particles to be expelled from the current channels, due to the magnetic deformations resulting from the tilt-kink. 
In terms of the evolving energy distribution, a representative case for 20000 electrons is shown in Fig.~\ref{f-B2}. The original thermal population (black dashed line) evolves with time as indicated by the colour legend: a clear power law tail develops. This case used a resistivity for the particle dynamics that is much lower than the $\eta_{\rm{MHD}}=10^{-4}$ used in the resistive MHD simulation. As there is no feedback of the particles to the electromagnetic fields, one can indeed use a different, more realistic (i.e. lower) resistivity in Ohm's law to quantify the parallel electric field for particle evolution, and hence get closer to true coronal conditions. The computation from Fig.~\ref{f-B2} also used the thermal bath technique to avoid unrealistic non-thermal tails to develop. This thermal bath approach at the same time accounts for the finite length conditions of line-tied loops, where energetic particles would be lost due to collisions with more dense chromospheric material. In the case shown in Fig.~\ref{f-B2}, the flux rope length adopted is 60 Mm, or about 0.086 $R_\odot$, and we get electron energies up to 4 MeV. Doubling the flux rope length effectively doubles the maximum attainable energy. For protons, energies up to 1 GeV could be explained by the tilt-kink and associated reconnection dynamics.

\section{Coalescence studies}

Closely related to the tilt instability is the coalescence process, where parallel current systems attract. This process is frequently studied in 2D (or 2.5D) setups~\cite{Biskamp1980} and appears naturally when a (reconnecting) current sheet forms multiple islands (e.g. by tearing), which then coalesce.  Several force-free MHD equilibria that consist of neighbouring flux tubes can likewise access a lower energy state configuration~\cite{Longcope1993} by coalescing parallel current channels. The resulting lower energy state shows pronounced current sheets. This is also a linear ideal MHD instability, but ultimately the reconnection within the current layers is expected to play a prominent role. In~\cite{Longcope1993}, the configuration used a checkerboard cross-sectional pattern of parallel and antiparallel current systems, and the combination of antiparallel repelling islands and parallel attracting flux systems was then analysed in a periodic 2D setup.  In a 3D variant of this force-free setup, the effects of line-tying were incorporated in (resistive) zero-$\beta$ simulations~\cite{Longcope1994}. Again, current layers develop naturally between similar helicity, attracting flux tubes. In what follows, we discuss related findings from recent numerical simulations, where present model efforts can attain more realistic magnetic Reynolds numbers, or even incorporate full kinetic aspects of the reconnection process. 

\subsection{Island coalescence}

\begin{figure*}
\centerline{\includegraphics[width=\textwidth]{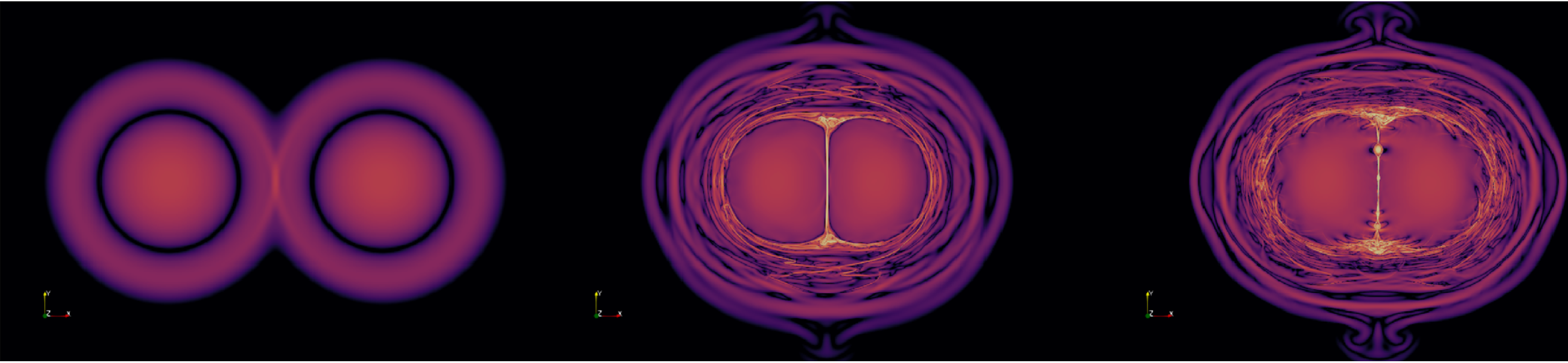}}
\caption{Island coalescence simulations from~\cite{Ripperda2019}. Shown is the out of plane current component for three representative runs: one where no perturbation is applied and hence the ideal equilibrium maintains its neighbouring island structure (left), and two where a small perturbation initiates the coalescence process (middle and right panel). Both latter simulations are in the far nonlinear regime where the central current sheet is fully resolved. That the rightmost case has tearing occuring in the central current sheet is fully consistent with its lower resistivity value of $\eta= 5\times 10^{-5}$. }\label{f-B3}
\end{figure*}

Figure~\ref{f-B3} shows the result of three different (actually, special relativistic) resistive 2.5D MHD evolutions, which start off from a pair of force-free cylindrical flux ropes~\cite{Ripperda2019}. All three panels show the instantaneous out-of-plane current density magnitude, and the leftmost one is virtually identical to the initial condition employed. A Lundquist-type force-free configuration~\cite{Lyutikov2017} realises a zero total current within each flux rope, by making a nested parallel-antiparallel current system: the current magnitude is seen to vanish not only outside each flux rope, but also on a specific radius for each rope (the out-of-plane magnetic field component does not change sign). This configuration of two adjacent flux ropes represents an MHD equilibrium. The leftmost panel confirms this: it represents a snapshot of an unperturbed simulation, which manages to maintain the adjacent flux rope structure. The right two panels are snapshots from initially perturbed cases, in the far nonlinear evolution of coalescence. Those two cases differ only in the value of the resistivity parameter: the middle case uses $\eta=10^{-4}$, while the rightmost case has $\eta=5 \times 10^{-5}$. It is seen that the coalescing flux ropes develop extremely narrow current sheets in between, and a threshold for the resistivity (or the Lundquist number) exists below which this sheet itself can become tearing unstable and develop island structures. This is indeed happening in the rightmost case shown in Fig.~\ref{f-B3}. These islands themselves can coalesce as well, and may dynamically get ejected out of the current sheet to enhance turbulent structure seen around the merged flux ropes. The results from~\cite{Ripperda2019} focused on varying the plasma-beta (between 0.01 and 1) and magnetization parameters, the latter being of specific interest for relativistic reconnection regimes. For solar contexts, it is noteworthy to recall that extreme effective resolutions (on the order of $8000^2$ or $16000^2$) were needed to justifiably claim numerically resolving the high magnetic Reynolds (or Lundquist) number regime. This also means that actual 3D simulations should exploit adaptive mesh refinement, or alternatively make use of spectral methods, to fully probe the reconnection process. The link between this ideal MHD coalescence process, and the merge scenario advocated in~\cite{Linton2001} for colliding flux tubes is at present poorly understood. In the same spirit, the combined effects of line-tying, flux rope curvature, and full thermodynamic effects in resistive MHD settings is a challenge for future simulations of interacting current systems. 

\subsection{Coupled MHD-kinetic approaches}

All studies discussed thus far employed pure MHD viewpoints on the solar coronal dynamics, but it is well known that the details of reconnection need a proper kinetic treatment to capture not only the scale separations associated with electron-ion dynamics, but also the intricate details of velocity phase space behaviour. An important step forward combines the merits of both MHD and kinetic approaches by establishing an embedded region within a larger scale MHD domain, which is treated with a particle in cell (PIC) simulation. The `Coupled MHD And PIC' or CMAP approach was demonstrated on a variety of 2D test problems in~\cite{Makwana2017}, refining on earlier efforts by~\cite{Daldorff2014}, where a similar two-way coupling of MHD and embedded PIC was termed MHD-EPIC. 

\begin{figure*}
\centerline{\includegraphics[width=0.49\textwidth]{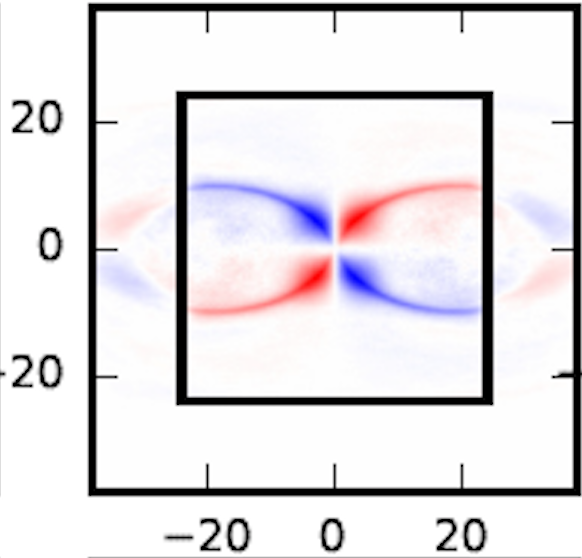}\includegraphics[width=0.49\textwidth]{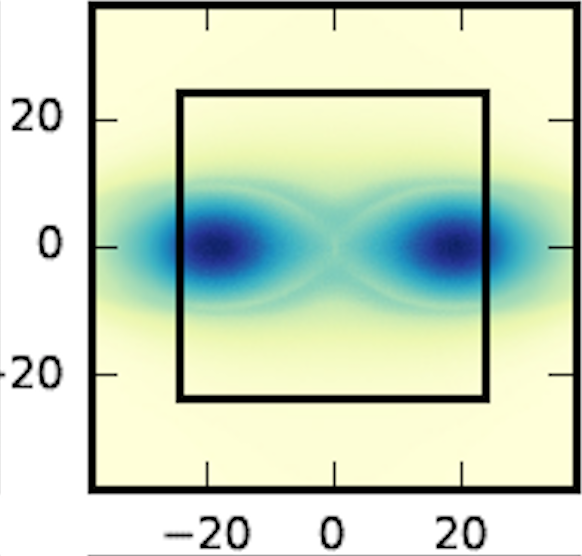}}
\caption{The out of plane magnetic field component (left) and the thermal pressure (right) in a coalescence simulation where MHD is coupled to an embedded PIC region (in the central square indicated). From~\cite{Makwana2018}, the length units are in ion skin depths.}\label{f-Kirit}
\end{figure*}
In a follow-up application, \cite{Makwana2018} investigated the coalescence process for a 2.5D setup, but where the initial magnetic field perpendicular to the simulated plane vanishes. Two magnetic islands are initially about 60 ion skin depths ($d_i$) apart from one another, but the coalescence process brings them closer together. Figure~\ref{f-Kirit} shows the central area of a simulation which has an MHD domain of size $120 d_i \times 120 d_i$. Using a static adaptive mesh, this central area is covered by the finest MHD grid level, and within this grid, an area of size $48 d_i \times 48 d_i$ has been initiated from the MHD with a PIC region (the central black square on both panels of Fig.~\ref{f-Kirit}). The CMAP approach manages a smooth transition between both representations across this coupling interface, which was improved from earlier works by means of a heuristic mixing of the MHD with the kinetic result using a weighted averaging. Shown are the pressure (right) and out of plane magnetic field component (left). The latter vanishes in a pure MHD setup, as the typical quadrupolar pattern is only realised as soon as Hall effects are incorporated, but here it is properly recovered in the CMAP model. The central region dynamics in the CMAP simulation was found to be very close to a reference simulation where the entire domain was treated with PIC. This could be quantified, by means of the obtained reconnection rate agreement. The true power of the CMAP approach is that it can handle much larger domain sizes than a PIC approach, and this was shown in coalescence studies of island structures covering $480 d_i \times 480 d_i$. Indeed, full PIC treatments of 2D coalescing setups~\cite{Du2018} begin to identify the various processes that play a role in particle acceleration and energization, and this in box sizes of $25-100 d_i$.

\subsection{Coalescing current systems: a novel cartoon view inspired by observations}

While the previously mentioned theoretical studies of coalescing flux tubes highlight the need for a proper treatment of the different dynamical scales (fluid to kinetic) involved, a recent observation of a coronal mass ejection~\cite{Gou2019} has been interpreted as forming a leading flux rope during its eruption process, continuously fed by mini flux ropes seen as plasmoids. This challenges the usual view of a pre-existing, MHD unstable flux rope according to the standard solar flare model. This particular event seemingly forms out of a slowly rising, upwardly extending current sheet, which fragments in multiple plasmoids with observed widths of 2 arseconds. Its later evolution is consistent with coalescing plasmoid behavior, and highlighted how a flux rope can be formed during eruption, in an intrinsically cross-scale manner.

\section{Outlook}

Ideal MHD instabilities that are responsible for much of the flux rope dynamics seen in the solar corona can be current-driven kink instability, the strapping field associated torus instability, or in the case of interacting flux ropes, the tilt, tilt-kink as well as coalescence instability. Most likely, mixtures of these and other MHD modes (such as of interchange type) cooperate to drive topological reconfigurations, and initiate eruptions. Reconnection details start to be resolved with unprecedented detail, revealing multiple interacting islands or substructure flux ropes within the current sheets underneath eruptions. These fine-scale details may well be confirmed by present and forthcoming solar observatories and space missions. The idealized setups for tilt-kink or coalescing flux tubes discussed must progress to setups inspired by observations, which do involve line-tying, and the details of how the ambient field that straps (single or multiple) current loops behave when reconnection occurs. Current model efforts to interpret flux rope activity vary in scope: some use vector magnetic field info to provide nonlinear force-free field (NLFFF) extrapolations to quantify stability tresholds, others take into account thermodynamic variations in fully compressible MHD studies. The latter efforts started to go beyond pure MHD effects, by studying particle acceleration aspects, or using coupled MHD and kinetic physics. The way forward will be to exploit all models in some data-driven way. As a promising example, NLFFF modeling of the source region of a complex CME~\cite{Awasthi2018}, not fully fitting the single flux rope scenario when observed further out in the heliosphere, found evidence for a multi-flux-rope system acting as its source region, characterized by reconnection and plasma flaring. A challenge for future (multi-scale) modeling efforts, is to use such NLFFF quantifications of multi-rope systems as initial conditions for full MHD evolutions.

In order to properly simulate the high magnetic Reynolds number regime, or accommodate for collisionless effects through coupled MHD-PIC simulations, modern open source software frameworks like MPI-AMRVAC~\cite{mpiamrvac2018,mpiamrvac2014,mpiamrvac2012} provide useful input to the scientific community. Indeed, several of our example applications~\cite{Xia2014,Zhaoetal2017,Zhaoetal2019,Mei2018,Keppens2014,Ripperda2017,Ripperda2019,Makwana2018} made use of MPI-AMRVAC. This framework can combine NLFFF quantifications~\cite{Guo2016a,Guo2016b} with follow-up MHD evolutions, and starts to exploit data driven boundary prescriptions~\cite{Guo2019}. Its flexible design can directly contrast zero-$\beta$ with finite-$\beta$ models, up to non-adiabatic MHD evolutions in stratified conditions, and thereby enhance our understanding of coronal eruptions.

\begin{acknowledgements}
RK thanks Chun Xia at Yunnan University, Kunming; Weiqun Gan at Purple Mountain Observatory, Nanjing; as well as Chen Peng-Fei at Nanjing University, Nanjing, for kind hospitality during his sabbatical stay. RK was supported by a joint FWO-NSFC grant G0E9619N and by his ERC Advanced Grant PROMINENT. BR was supported by an Alexander von Humboldt Fellowship. YG was supported by NSFC (11773016, 11733003, 11533005 and 11961131002). This project has received funding from the European Research Council (ERC) under the European Union’s Horizon 2020 research and innovation programme (grant agreement No. 833251 PROMINENT ERC-ADG 2018).
\end{acknowledgements}

On behalf of all authors, the corresponding author states that there is no conflict of interest.
% BibTeX users please use one of
%\bibliographystyle{spbasic}      % basic style, author-year citations
\bibliographystyle{spmpsci}      % mathematics and physical sciences
%\bibliographystyle{spphys}       % APS-like style for physics
%\bibliography{rmpp}   % name your BibTeX data base

% Non-BibTeX users please use
%\begin{thebibliography}{}
%
% and use \bibitem to create references. Consult the Instructions
% for authors for reference list style.
%
%\bibitem{RefJ}
% Format for Journal Reference
%Author, Article title, Journal, Volume, page numbers (year)
% Format for books
%\bibitem{RefB}
%Author, Book title, page numbers. Publisher, place (year)
% etc
%\end{thebibliography}

\end{document}